\documentclass[aip, reprint]{revtex4-1}
\usepackage{amsmath, amsthm, amssymb, graphicx, epstopdf, dcolumn}
\begin{document}
\title{Femtosecond X-ray magnetic circular dichroism absorption spectroscopy at
an X-ray free electron laser}
\author{Daniel J. Higley}
\email{dhigley@stanford.edu}
\affiliation{SLAC National Accelerator Laboratory, 2575 Sand Hill Road, Menlo
Park, California 94025, USA}
\affiliation{Department of Applied Physics, Stanford University,
Stanford, California 94305, USA}
\author{Konstantin Hirsch}
\affiliation{SLAC National Accelerator Laboratory, 2575 Sand Hill Road, Menlo
Park, California 94025, USA}
\author{Georgi L. Dakovski}
\affiliation{SLAC National Accelerator Laboratory, 2575 Sand Hill Road, Menlo
Park, California 94025, USA}
\author{Emmanuelle Jal}
\affiliation{SLAC National Accelerator Laboratory, 2575 Sand Hill Road, Menlo
Park, California 94025, USA}
\author{Edwin Yuan}
\affiliation{SLAC National Accelerator Laboratory, 2575 Sand Hill Road, Menlo
Park, California 94025, USA}
\affiliation{Department of Applied Physics, Stanford University,
Stanford, California 94305, USA}
\author{Tianmin Liu}
\affiliation{SLAC National Accelerator Laboratory, 2575 Sand Hill Road, Menlo
Park, California 94025, USA}
\affiliation{Department of Physics, Stanford University,
Stanford, California 94305, USA}
\author{Alberto A. Lutman}
\affiliation{SLAC National Accelerator Laboratory, 2575 Sand Hill Road, Menlo
Park, California 94025, USA}
\author{James P. MacArthur}
\affiliation{SLAC National Accelerator Laboratory, 2575 Sand Hill Road, Menlo
Park, California 94025, USA}
\affiliation{Department of Physics, Stanford University,
Stanford, California 94305, USA}
\author{Elke Arenholz}
\affiliation{Lawrence Berkeley National Laboratory, Berkeley, California
94720, USA}
\author{Zhao Chen}
\affiliation{SLAC National Accelerator Laboratory, 2575 Sand Hill Road, Menlo
Park, California 94025, USA}
\affiliation{Department of Physics, Stanford University,
Stanford, California 94305, USA}
\author{Giacomo Coslovich}
\affiliation{SLAC National Accelerator Laboratory, 2575 Sand Hill Road, Menlo
Park, California 94025, USA}
\author{Peter Denes}
\affiliation{Lawrence Berkeley National Laboratory, Berkeley, California
94720, USA}
\author{Patrick W. Granitzka}
\affiliation{SLAC National Accelerator Laboratory, 2575 Sand Hill Road, Menlo
Park, California 94025, USA}
\affiliation{Van der Waals-Zeeman Institute, University of Amsterdam, 1018XE
Amsterdam, The Netherlands}
\author{Philip Hart}
\affiliation{SLAC National Accelerator Laboratory, 2575 Sand Hill Road, Menlo
Park, California 94025, USA}
\author{Matthias C. Hoffmann}
\affiliation{SLAC National Accelerator Laboratory, 2575 Sand Hill Road, Menlo
Park, California 94025, USA}
\author{John Joseph}
\affiliation{Lawrence Berkeley National Laboratory, Berkeley, California
94720, USA}
\author{Lo{\"i}c Le Guyader}
\affiliation{Institute for Molecules and Materials (IMM), Radboud University
Nijmegen, Heyendaalseweg 135, 6525 Aj Nijmegen, The Netherlands}
\affiliation{European XFEL GmbH, Albert-Einstein-Ring 19, 22761
Hamburg, Germany}
\author{Ankush Mitra}
\affiliation{SLAC National Accelerator Laboratory, 2575 Sand Hill Road, Menlo
Park, California 94025, USA}
\author{Stefan Moeller}
\affiliation{SLAC National Accelerator Laboratory, 2575 Sand Hill Road, Menlo
Park, California 94025, USA}
\author{Hendrik Ohldag}
\affiliation{SLAC National Accelerator Laboratory, 2575 Sand Hill Road, Menlo
Park, California 94025, USA}
\author{Matthew Seaberg}
\affiliation{SLAC National Accelerator Laboratory, 2575 Sand Hill Road, Menlo
Park, California 94025, USA}
\author{Padraic Shafer}
\affiliation{Lawrence Berkeley National Laboratory, Berkeley, California
94720, USA}
\author{Joachim St{\"o}hr}
\affiliation{SLAC National Accelerator Laboratory, 2575 Sand Hill Road, Menlo
Park, California 94025, USA}
\author{Arata Tsukamoto}
\affiliation{Department of Electronics and Computer Science, Nihon University,
7-24-1 Narashino-dai Funabashi, Chiba 274-8501, Japan}
\author{Heinz-Dieter Nuhn}
\affiliation{SLAC National Accelerator Laboratory, 2575 Sand Hill Road, Menlo
Park, California 94025, USA}
\author{Alex H. Reid}
\affiliation{SLAC National Accelerator Laboratory, 2575 Sand Hill Road, Menlo
Park, California 94025, USA}
\author{Hermann A. D{\"u}rr}
\affiliation{SLAC National Accelerator Laboratory, 2575 Sand Hill Road, Menlo
Park, California 94025, USA}
\author{William F. Schlotter}
\affiliation{SLAC National Accelerator Laboratory, 2575 Sand Hill Road, Menlo
Park, California 94025, USA}
\begin{abstract}
X-ray magnetic circular dichroism spectroscopy using an X-ray free electron laser is demonstrated with spectra over the Fe L$_{3,2}$-edges. This new ultrafast time-resolved capability is then applied to a fluence-dependent study of all-optical magnetic switching dynamics of Fe and Gd magnetic sublattices in a GdFeCo thin film above its magnetization compensation temperature. At the magnetic switching fluence, we corroborate the existence of a transient ferromagnetic-like state. The timescales of the dynamics, however, are longer than previously observed below the magnetization compensation temperature. Above and below the switching fluence range, we observe secondary demagnetization with about 5 ps timescales. This indicates that the spin thermalization takes longer than 5 ps.
\end{abstract}
\maketitle


X-ray Absorption Spectroscopy (XAS) is one of the most commonly practiced techniques at synchrotron light sources. Its ability to probe local electronic structure with elemental specificity makes XAS indispensable in diverse fields such as chemistry, materials science, and magnetism.\cite{stohr2013nexafs, de2008core, stohr2006magnetism} The soft X-ray range is of great importance for many studies due to the accessibility of the K-edges of O, N, and C, as well as the 3d transition metal L-edges, and rare earth M-edges.

Polarization resolution is critical in many XAS studies. Examples include determination of the orientation of molecules on surfaces,\cite{stohr2013nexafs} and the detailed orbital and spin structure of strongly correlated materials.\cite{aetukuri2013control} In addition, X-Ray Magnetic Circular Dichroism spectroscopy (XMCD) has proven to be of tremendous value in studying magnetism down to the nanometer length-scale.\cite{stohr2006magnetism}

Extension of soft X-ray XAS to the femtosecond timescale offers the potential to follow and ultimately disentangle chemical and materials processes on their natural timescales.\cite{dell2013real, cavalleri2005band}. ``Femtoslicing'' synchrotron beamlines, which are capable of delivering $\sim$100 fs X-ray pulses, albeit at low intensity, have begun the realization of this potential with many important results.\cite{stamm2007femtosecond, radu2011transient, bergeard2014ultrafast, huse2009probing} However, the long data acquisition times at femtoslicing sources means systematic studies, as well as those requiring very high accuracy are prohibitively time consuming. X-ray Free Electron Lasers (XFELs)\cite{mcneil2010x, emma2010first} provide femtosecond X-ray pulses with spectacularly high brightness and offer the potential to solve this problem. The strenghts of XFELs have already been used in tremendously successful Imaging and scattering studies.\cite{chapman2011femtosecond, wang2012femtosecond, graves2013nanoscale} Femtosecond XAS in the soft X-ray range, however, has been more challenging. High pulse energies, strong pulse-to-pulse fluctuations,\cite{saldin2010statistical} and low repetition rates at XFELs hinder implementing spectroscopy in the same ways as have been perfected at synchrotron light sources over decades.  Nonetheless, several important studies demonstrated the potential of XAS in the soft X-ray range at XFELs,\cite{bernstein2009near} especially using partial fluorescence yield detection\cite{dell2013real, ostrom2015probing}. The direct and quantitative method of detection of incident and transmitted monochromatic X-rays, however, remained difficult.


\begin{figure*}
\centerline{\includegraphics[width=17cm]{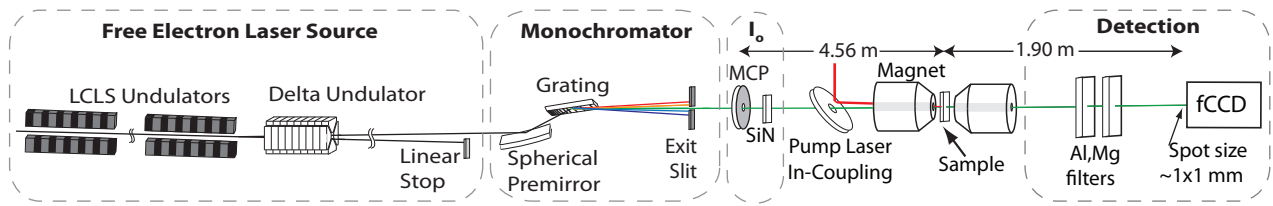}}
\caption{Experimental setup. The Delta undulator produced circularly polarized X-rays which are then monochromatized. The total energy of these X-rays is detected with the fluorescence-based I$_0$ detector before they are transmitted through the sample and detected with a system of attenuating filters and the fCCD.}
\label{fig:setup}
\end{figure*}

We overcome these challenges using a new measurement setup with robust, high precision detection of incident and transmitted X-rays. The reliable detection of transmitted X-rays had been particularly difficult in the past and was overcome by using a high linearity CCD with a flexible X-ray attenuation system. Together with the newly installed variable polarization ``Delta'' undulator,\cite{lutman2015polarization} this setup enabled time-resolved XMCD studies of unprecedented accuracy. The transmission of monochromatic X-rays was measured to an accuracy of 1\% of the transmission within 1 s of measurement, and better accuracy is obtainable through averaging further. This new capability is demonstrated with comparison of static XMCD spectra over the Fe L-edges of a GdFeCo thin film recorded at the XFEL and that recorded at a synchrotron light source. We then apply this technique to a fluence-dependent study of the Gd and Fe magnetic sublattice dynamics in GdFeCo during all-optical switching.

The measurement setup (Fig. \ref{fig:setup}) was implemented at the Soft X-Ray Materials Science Instrument\cite{schlotter2012soft, heimann2011linac} (SXR) of the Linac Coherent Light Source\cite{emma2010first} (LCLS) XFEL. The Delta undulator operated in the diverted beam scheme, producing circularly polarized X-ray pulses of $\sim$200 $\mu$J pulse energy and 25 fs FWHM duration.\cite{lutman2015polarization} A grating monochromator with 100 lines/mm then filtered these X-rays to a bandwidth of about 200 meV.\cite{heimann2011linac} Following the monochromator, a novel incident X-ray flux (I$_0$) detector, described below, measured the incoming X-ray pulse energy. The X-ray beam then traversed a pair of Kirkpatrick-Baez mirrors, giving a 50 $\mu$m FWHM diameter X-ray spot at the sample. The X-ray probe and an optical pump laser were normally incident on a sample with a variable time delay between them. The pump laser had a wavelength of 800 nm, a diameter of 200 $\mu$m FWHM at the sample, a duration of 60 fs FWHM, and linear polarization. The sample was magnetized using an electromagnet with an applied field of +/- 200 mT. The X-rays transmitted through the sample propagated through a 200 nm aluminum film to separate them from the optical pump laser beam. Finally, these X-rays were attenuated and then detected with a CCD, as described further below.

A key enabler for these experiments was the development of new detectors for incident and transmitted X-rays. The I$_0$ detector enabled detection of incident X-rays with a higher sensitivity than previously possible,\cite{tiedtke2014absolute} allowing high precision experiments with the lower photon throughput of the Delta undulator ($\sim$200 $\mu$J pulses, in contrast to $\sim$1 mJ pulses in standard operation). The detector consists of a Micro-Channel Plate (MCP) assembly of two MCPs and a single metal anode (Hamamatsu F2223-21SH). A 4 mm diameter hole in the center of the assembly let the X-ray beam pass through to a 300 nm thick Si$_3$N$_4$ membrane window placed 2.5 cm downstream of the MCP.  On passing through the Si$_3$N$_4$ membrane, fluorescent X-rays were emitted, some of which were incident on the 5.7 cm$^2$ active area of the MCP. The MCP was biased at about -1400 V and the signal was sent through a low noise preamplifier and low-pass filter (SR570) before being digitized. The signal from each pulse for this detector was taken as the integral over the narrow temporal window of the electronic trace due to the X-ray pulse.

To measure the transmitted X-rays, we used the cooled, in-vacuum fast CCD (fCCD)\cite{doering20121mpixel} and attenuating filters (Al and Mg filters of Fig. \ref{fig:setup}). The signal-to-noise ratio of this detector is ultimately limited by the number of X-ray photons which can be detected without reaching pixel saturation. To maximize this number, the fCCD was placed as far downstream from the sample as was practical. This allowed the beam to spread out to 1 mm FWHM diameter at the fCCD. In addition, the filter system was designed to give fine steps of attenuation over a large range: steps of factors of three to over 10$^5$ near the Fe L$_3$ (707 eV) and Gd M$_5$ (1190 eV) edges. This allowed near optimal attenuation for any experimental situation.

\begin{figure}
\centerline{\includegraphics[width=8.5cm]{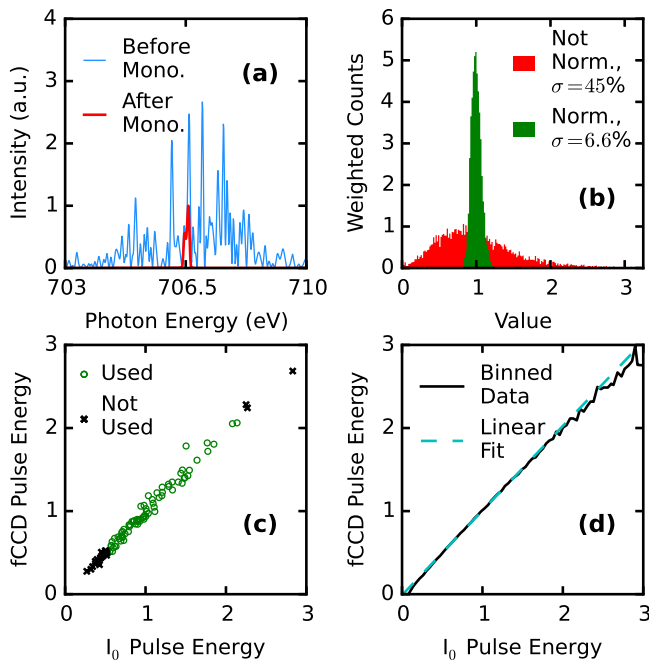}}
\caption{Performance of X-ray absorption measurement system. (a) Calculated energy spectral density of an XFEL pulse due to the self-amplified spontaneous emission process (light blue).\cite{saldin2010statistical} The intensity transmitted through a 150 meV monochromator is shown in red. (b) Histograms of the pulse energy after the monochromator as measured by the fCCD (red), and by the fCCD normalized by the I$_0$ detector (green). (c) Pulse energy of monochromatic XFEL pulses detected at the fCCD vs I$_0$ detector. Pulses of low or very high energy are not used in data analysis. (d) Average fCCD measured pulse energy versus I$_0$ pulse energy bin.}
\label{fig:noise}
\end{figure}

The performance of this experimental setup is characterized in Fig. \ref{fig:noise}. The origin of large fluctuations in pulse energy is illustrated in Fig. \ref{fig:noise}a (where the pulse energy we refer to is the summed energy of all photons in the pulse, not the central photon energy). The self-amplified spontaneous emission process\cite{saldin2010statistical} gives a ``spiky'' spectrum (light blue), which, when filtered with a narrow bandwidth monochromator for spectroscopy (red), results in large fluctuations of the pulse energy. The distribution in pulse energies, as measured by the fCCD after the monochromator, is shown by the wide distribution of the red histogram in Fig. \ref{fig:noise}b, with a standard deviation of 45\% of the mean. When normalized by the pulse energy measured by the I$_0$ detector, this standard deviation is reduced to 6.6\% of the mean, as shown by the narrow green histogram in Fig. \ref{fig:noise}b. Furthermore, the standard deviation of the normalized transmitted X-ray pulse energies was found to decrease inversely proportionally to the square root of the number of X-ray pulses averaged, down to a standard deviation of at most 0.1\% of the mean. For optimum linearity and signal-to-noise ratio of the measurement system, it is critical to tune the attenuation in front of the fCCD as well as the MCP bias. For this, correlation graphs such as those shown in Fig. \ref{fig:noise}c and \ref{fig:noise}d were optimized in real time during the experiments.  Even with optimization, some fraction of the pulses with the highest or lowest pulse energies were typically discarded due to being outside the optimal pulse energy range (in this case, those below the 20th percentile or above the 95th percentile of pulse energy).


\begin{figure}
\centerline{\includegraphics[width=7.62cm]{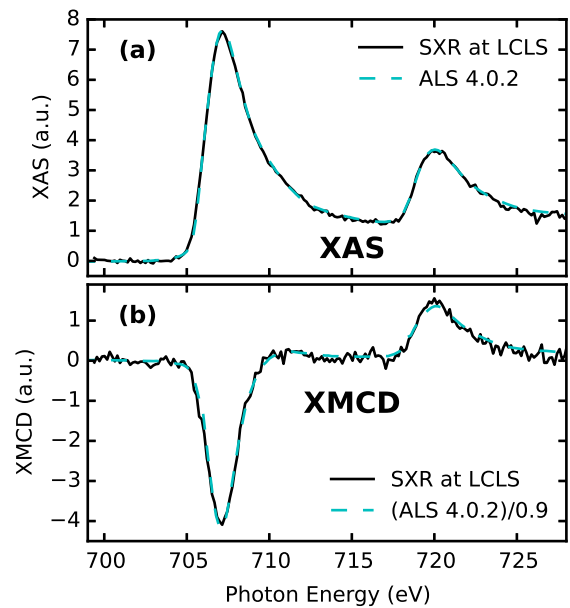}}
\caption{Comparison of static Fe L-Edge spectra recorded on a 40 nm GdFeCo sample at LCLS and beamline 4.0.2 of the ALS synchrotron light source. (a)  XAS as would be measured with linear polarization, found by taking the average of XAS with parallel and anti-parallel alignment of X-ray helicity and sample magnetization. (b) XMCD, found by taking the difference of XAS with parallel and anti-parallel alignment of X-ray helicity and sample magnetization and dividing by the degree of polarization of the source. The LCLS degree of circular polarization was 100 percent within the accuracy of its measurement, so the XMCD measured by LCLS is not divided by a factor to account for an imperfect source polarization.}
\label{fig:spec}
\end{figure}

With this new measurement setup, we recorded XMCD spectra over the Fe L$_{3,2}$-edges and compared the results to those recorded on the same sample at a synchrotron light source, beamline 4.0.2 of the Advanced Light Source (ALS) (Fig. \ref{fig:spec}). The sample was an amorphous, out-of-plane magnetized, 40 nm thick Gd$_{24}$Fe$_{66.5}$Co$_{9.5}$ thin film characterized elsewhere.\cite{graves2013nanoscale} It was prepared via magnetron sputtering on a 100 nm Si$_3$N$_4$ membrane with a 5 nm Si$_3$N$_4$ buffer layer and 10 nm capping layer. The XAS and XMCD spectra were obtained from measurements with opposite magnetizations at LCLS while opposite X-ray helicities were used at ALS. For amorphous samples such as here, these measurements should give identical results.\cite{stohr1995determination} A linear background fitted to the pre-edge was subtracted from each XAS spectrum which were then normalized to have equal L$_3$ intensity, and the XMCD spectra were scaled by the same factor. The XMCD spectra collected at ALS were also divided by the source polarization of 90\%.\cite{young2001first} The XAS in (a) agree with an RMS deviation of 1.1\% of the peak magnitude, while the XMCD in (b) have an RMS deviation of 3.1\% of the peak magnitude.

From this data, we also determined the degree of circular polarization provided by the Delta undulator at LCLS. To do so, we compared the integral over the Fe L$_3$ XMCD signal (704 to 711 eV) at LCLS and ALS. Then, accounting for the known 90\% degree of polarization provided by the ALS, the LCLS degree of circular polarization was found to be 98+2/-4\% using three spectra like the one shown in Fig. \ref{fig:spec}.

\begin{figure}[htb]
\centerline{\includegraphics[width=8.5cm]{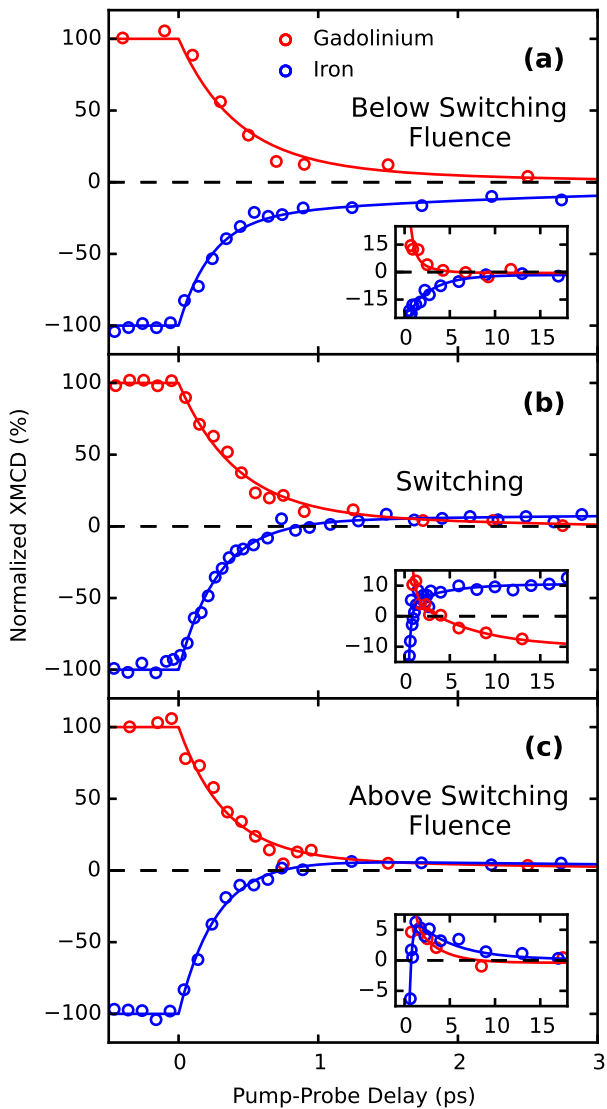}}
\caption{Ultrafast dynamics of the Fe and Gd sublattice magnetizations after optical excitation and near the switching threshold, measured through the XMCD signal at the Fe L$_3$ and Gd M$_5$ edges. (a) 20 mJ/cm$^2$, below the switching threshold fluence. (b) 25 mJ/cm$^2$, within the switching fluence range. (c) 35 mJ/cm$^2$, above the switching fluence range. The circles indicate measured data points, with the solid lines being fits with two exponential decay times to the data.}
\label{fig:dynamic}
\end{figure}

With the improved sensitivity of our measurement setup and circular polarization capability, it is now possible to use the superior flux of XFELs to systematically investigate ultrafast element-specific magnetization dynamics. A prime example is all-optical switching, where the sole action of a femtosecond laser pulse triggers, in certain materials, a deterministic reversal of the magnetization.\cite{stanciu2007all, radu2011transient, graves2013nanoscale, kirilyuk2013laser, lambert2014all, mangin2014engineered} A seminal result obtained at a soft X-ray femtoslicing beamline revealed that in the most studied of these materials, GdFeCo alloys, this occurs through a highly non-equilibrium transient ferromagnetic alignment of the otherwise antiferromagnetically coupled Fe and Gd sub-lattice magnetizations.\cite{radu2011transient} However, the long data acquisition times at femtoslicing sources have so far prevented more throrough studies with varying sample and excitation parameters. In a step towards remedying this, we study the fluence-dependence of all-optical switching of GdFeCo above its magnetization compensation temperature, in contrast to previous studies.\cite{radu2011transient} We note that many equilibrium and non-equilibrium properties have different behaviors above and below the magnetization compensation temperature.\cite{guyader2014all, medapalli2012efficiency, kirilyuk2013laser}

The sample was a 20 nm thick Gd$_{24}$Fe$_{66.5}$Co$_{9.5}$ thin film. Measurements were performed at 300 K, which is above the sample's magnetization compensation temperature of 227 K. The XMCD signal was recorded at fixed photon energy and X-ray polarization while the sample magnetization was switched every 30 s during data acquisition. The temporal evolution of the Fe 3d and Gd 4f magnetizations as probed by XMCD at the Fe L$_3$ and Gd M$_5$ edges are shown in Fig. \ref{fig:dynamic} for three different fluences: (a) below, (b) within, and (c) above the all-optical switching fluence window. The temporal resolution of these measurements was determined by the X-ray arrival time jitter of about 100 fs. Following Radu et al.,\cite{radu2011transient} we fitted the femtosecond demagnetization and picosecond switching behavior to a function with two exponentials. The Fe magnetization dynamics had time constants of 250 fs and 4 ps, whereas that of Gd was 400 fs and 6 ps. Within the measurement error, these time constants did not evolve significantly over the fluence range shown.

Below the switching threshold fluence (Fig. \ref{fig:dynamic}a), we simply observe demagnetization on both a fast femtosecond and slower picosecond timescale. At the all-optical switching fluence (Fig. \ref{fig:dynamic}b), we observe the typical behavior previously reported.\cite{radu2011transient} Initially, the Fe and Gd magnetizations decay with typical demagnetization timescales.\cite{koopmans2010explaining} Fe demagnetizes faster than Gd and reverses magnetization at 870 fs. This leads to a transient ferromagnetic state that exists until 3.7 ps, when the Gd magnetization also reverses. We note that these timescales are significantly longer than what has been reported before. \cite{radu2011transient} This may be due to the fact that here the starting temperature was set to room temperature, above the magnetization compensation temperature of 227 K, while Radu et al. \cite{radu2011transient} reported on switching below the magnetization compensation temperature.

Interestingly, we still see switching in the Fe magnetic sublattice far above the switching fluence while the Gd sublattice only shows demagnetization (Fig. \ref{fig:dynamic}c). This also leads to ferromagnetic alignment of Fe and Gd which, however, is found to persist to longer times. In the long time behavior shown in the inset of Fig. \ref{fig:dynamic}c, we see evidence for a second demagnetization timescale of about 5 ps on which both Fe and Gd magnetization is quenched. This behavior and the two timescales of demagnetization found below the switching threshold could indicate that following the initial sub-ps demagnetization the system is still in a non-equilibrium state and true spin thermalization can take more than 5 ps.


To summarize, XFEL and beamline instrumentation developments at LCLS have enabled robust ultrafast, polarization-resolved X-ray absorption spectroscopy in the soft X-ray range. Using this capability, we observed all-optical switching with ultrafast XMCD at an XFEL for the first time. In the future, the unique ability to take high resolution polarization-resolved XAS with ultrafast time resolution over entire edges will be invaluable in a number of fields. For example, in ultrafast magnetism studies, XMCD and XAS spectra are both expected to have significant,  but as yet unobserved, photon energy-dependent changes on sub-ps timescales.\cite{carva2009influence}

We would like to thank S. Carron, C. Ford, G. McCracken and C.P. O'Grady for technical assistance. This work is supported by the Department of Energy, Office of Science, Basic Energy Sciences, Materials Science and Engineering Division, under Contract No. DE-AC02-76SF00515, and the European Research Council (grant 339813 EXCHANGE). Use of the Linac Coherent Light Source, SLAC National Accelerator Laboratory, is supported by the U.S. Department of Energy, Office of Science, Office of Basic Energy Sciences under Contract No. DE-AC02-76SF00515. The Advanced Light Source is supported by the Director, Office of Science, Office of Basic Energy Sciences, of the U.S. Department of Energy under Contract No. DE-AC02-05CH11231. KH thanks the AvH foundation for financial support through the Feodor-Lynen program.
\bibliography{lcls_xmcd_refs}
\end{document}